\documentclass{aa}  

\usepackage{graphicx}
\usepackage{txfonts}
\usepackage{xcolor}
\usepackage{cprotect}
\usepackage{longtable}
 \usepackage{multicol}
\usepackage{supertabular}
\usepackage[breaklinks]{hyperref}
\bibpunct{(}{)}{;}{a}{}{,}             
\definecolor{cobalt}{rgb}{0.06, 0.2, 0.65}
\hypersetup{
  colorlinks,
  citecolor=cobalt,
  linkcolor=[rgb]{0.8, 0.2, 1.0},
  urlcolor=cobalt,
}
\newcommand\obsc[0]{\text{obsc}}
\begin{document}

   \title{Secrets in the shadow: High precision stellar abundances of fast-rotating A-type exoplanet host stars through transit spectroscopy}
   \titlerunning{High precision stellar abundances of fast-rotating A-type exoplanet host stars}

   \author{M. B. Lam
          \inst{1}
          \and H. J. Hoeijmakers \inst{1}
          \and B. Prinoth \inst{1,2}
          \and B. Thorsbro \inst{3,1}
          }

    \institute{Lund Observatory, Division of Astrophysics, Department of Physics, Lund University, Box 43, 221 00 Lund, Sweden \\
    \email{ma7177la-s@student.lu.se, jens.hoeijmakers@fysik.lu.se}
    \and
    European Southern Observatory, Alonso de Córdova 3107, Vitacura, Región Metropolitana, Chile
    \and
    Observatoire de la Côte d’Azur, CNRS UMR 7293, BP4229, Laboratoire Lagrange, F-06304 Nice Cedex 4, France
             }

   \date{}

 
  \abstract
   {The spectra of fast-rotating A-type stars have strongly broadened absorption lines. This effect causes blending of the absorption lines, hindering the measurement of the abundances of the elements that are in the stellar photosphere. }
   {As the exoplanet transits across its host star, it obscures the stellar spectrum that is emitted from directly behind the planet. We aim to extract this obscured spectrum because it is less affected by rotational broadening, resolving the blending of weak lines of elements that would otherwise remain inaccessible. This allows us to more precisely measure the metal abundances in ultra-hot Jupiter systems, many of which have fast rotating host stars.}
   {We develop a novel method that isolates the stellar spectra behind the planet during a spectral time-series, and reconstructs the disc-integrated non-broadened spectrum of the host star. We have systematically tested this method with model-generated spectra of the transit of WASP-189 b across its fast-rotating A-type host star, assessing the effects of limb darkening, choice of absorption lines, signal to noise regime; and demonstrating the sensitivity to photospheric parameters ($T_{\text{eff}}$, log $g$) and elemental abundances. We apply the method to observations by the HARPS high-resolution spectrograph.}
   {For WASP-189, we obtain the metallicity and photospheric abundances for several species previously not reported in literature, Mg, Ca and Ti, with significantly improved accuracy compared to the ordinary broadened stellar spectrum. This method can be generally applied to other transiting systems in which abundance determinations via spectral synthesis are imprecise due to severe line blending. It is important to accurately determine the photospheric properties of exoplanet host stars, as it can provide further insight into the formation and evolution of the planets.}
   {}


   \maketitle
%

\section{Introduction}

Physical parameters of exoplanets are often determined from observed properties of their host stars, including planetary radius, mass, temperature, age and orbital configuration \citep[e.g.][]{Winn2010, Buchhave2014}. Of particular interest is the star's chemical composition and how it compares to the chemical composition of the planet. This provides insight into the planetary formation process \citep[e.g.][]{Santos2004,Unterborn2017,Lothringer2021,Schulze2021}. However, early-type stars often have high projected rotational velocities, causing rotational broadening in their spectra that hinders the determination of photospheric abundances.

A-type main sequence stars have temperatures in the range of 7 500-10 000 K. As these main sequence stars form, angular momentum is conserved as the accretion disc collapses. As a result, more massive main sequence stars generally have high rotational velocities. From statistical surveys on projected stellar rotational velocities \citep[e.g.][]{Glebocki2005}, the stellar population can be divided into fast rotators with $v\sin i \gtrapprox 100$ km/s and slow rotators with $v\sin i \lessapprox 50$ km/s. Rapid rotation introduces line broadening \citep{1987ApJS...65..581G} and weak lines are smeared out. This limits the precise measurement of stellar properties from the stellar spectrum. A-type stars have traditionally been understudied with a focus on FGKM stars \citep[e.g.][]{Fischer2005,Cumming2008}.

Few studies exist on the abundances of A-type stars \citep[e.g.][]{Takeda2009}, however there is a bias towards metallic-line (Am) and chemically peculiar (Ap) A-type stars, which have under-/over-abundances of certain metallic elements (e.g. Ca, Sc, Sr), and that tend to be slow-rotating \citep[e.g.][]{Adelman_1973,1974ARA&A..12..257P,Smith1996}, or other slow-rotating A-type stars \citep{Royer2014}. However, the spectra of normal A-type stars are different from those of peculiar A stars due to diffusion in slow rotators \citep{Abt_2000}. Research has also been conducted on evolved A-type stars as they are slower rotating than those on the main sequence \citep[e.g.][]{Johnson2007,Bowler2010,Johnson2011,Sato2012}. Additionally, there exists research on A-type stars that are viewed pole-on, such as Vega \citep[e.g.][]{Gigas1986,Adelman1990}. Although Vega is a rapid rotator with a $v_{\text{eq}} \approx 160$ km/s, it has a $v\sin i_\star =21.9$ km/s \citep{2004IAUS..224...35H}, reducing the debilitating effect of line-broadening.

Despite being difficult to detect with radial velocity (RV) surveys \citep[e.g.][]{Galland2005,Borgniet2017}, over 20 A-type stars are known to host hot Jupiters to date \citep[e.g.][]{CollierCameron2010b,Hartman2015,Gaudi2017}. This paper proposes a new method that uses exoplanet transit time-series observations to isolate the non-broadened stellar spectrum by comparing the in-transit spectra in which part of the stellar disc is obscured, with the out-of-transit spectrum. Standard spectral synthesis methods can then be used to more precisely measure the photospheric properties, despite the high value of the projected velocity $v \sin i_\star$. This provides access to significantly improved chemical abundance determinations for this class of exoplanet host star. \\
\\
Using high-precision transit spectroscopy, the spectrum of the photosphere that is locally obscured by the transiting planet can be resolved \citep{Cegla_2016,Dravins2017}. This technique has been referred to as Doppler tomography \citep{Albrecht2007,CollierCameron2010}. The obscured stellar spectrum appears as an instability in the line shape during transit, which is also referred to as a Doppler shadow in studies that target the planet's absorption spectrum \citep[e.g.][]{Gaudi2017,Hoeijmakers2020}. In this paper, the Doppler shadow is defined as the narrow non-broadened stellar spectrum occulted by the transiting planet. This line deformation is also the cause of the Rossiter-Mclaughlin (RM) effect \citep[][]{1924ApJ....60...15R,1924ApJ....60...22M}, where it induces an apparent radial velocity (RV) shift as the planet traverses the stellar disc. The regions on the stellar disc that the planet obscures during the transit depend on the projected stellar equatorial velocity $v\sin i_\star$, the impact parameter of the planet $b$ (equivalent to the orbital inclination $i_p$), the misalignment between the planetary orbital axis and the stellar spin axis $\lambda$ and the relative size of the planet $R_p/R_\star$. Measuring the orbital misalignment via the RM effect is particularly important in the context of planet formation and evolution theories \citep{CollierCameron2010,Winn2010b,Albrecht2012}. \citet{Cegla_2016} built upon this method to account for differential stellar rotation, which was previously ignored. The tomography signal is particularly prominent for transits across fast-rotating host stars, where it has been used to verify the planetary nature of transit candidates \citep[e.g.][]{Gaudi2017,Anderson2018}.\\
\\
\noindent In this study, we adapted the Doppler tomography method to measure photospheric elemental abundances directly from the obscured stellar spectrum (and by-passing the cross-correlation operation) with a significantly increased accuracy than is possible in the strongly broadened and blended out-of-transit spectrum of the A-type host star WASP-189.

\section{Methods}

\subsection{Spectral synthesis} 
\label{sec:spectralsynthesis}

In this paper, we aim to model the effect of a transiting planet on the spectrum of the host star, and we use external software packages to model the stellar spectrum and carry out disc-integration. We used \verb|PySME| \citep{Valenti1996,Piskunov2017,pySME} to generate a non-rotation broadened stellar spectrum given the input stellar parameters:
   \[
      \begin{array}{lp{0.8\linewidth}}
         T_{\text{eff}}             & effective temperature in K             \\
         $[Fe/H]$          & metallicity in dex     \\
         \log g             & surface gravity in log(cgs)\\
         \mu & limb angle to evaluate spectra \\
         $[X/H]$ & elemental abundances in dex
      \end{array}
   \]
\verb|PySME| interpolates a model of the stellar atmosphere from a grid. The \textsc{ATLAS12} grid \citep{1993ASPC...44...87K} was used to model A-type stars in this paper. This grid covers a temperature range of 3500-50 000 K, $\log g$ between 0 -- 5 in steps of 0.5 and metallicity between -5 -- 1 dex in varying steps: +1.0, +0.5, +0.3, +0.2, +0.1, +0.0, -0.1, -0.2, -0.3, -0.5, -1.0, -1.5, -2.0, -2.5,-3.0, -3.5, -4.0, -4.5, and -5.0. The ATLAS12 grid covers the temperature range in uneven steps, and for the temperature range 3500 -- 10 000 K which we are interested in to model A-type stars, the step size is 250 K.

The VALD database \citep{1995A&AS..112..525P,2015PhyS...90e4005R} provided the input line list required for model synthesis and fitting. \verb|PySME| was configured to return a non-continuum normalised spectrum at a specified $\mu$ angle. We have also set macroturbulence and microturbulence velocities to 0, and applied no convolution to account for instrumental spectroscopic resolving power, so that all broadening of the spectra is solely due to rotation.

We used \verb|StarRotator|\footnote{https://github.com/Hoeijmakers/StarRotator} to simulate rotation broadening and line distortion during an exoplanet transit event, given the non-rotation broadened synthetic spectrum as an input. \verb|StarRotator| uses a grid-based approach to integrate over the stellar disc and computes the rotation broadened spectrum with the additional parameters
   \[
      \begin{array}{lp{0.8\linewidth}}
         v_{\text{eq}}  & equatorial velocity in m/s     \\
         i_\star               & stellar inclination in deg                    \\
         & number of $\mu$ angles
      \end{array}
   \]
Centre-to-limb variation was taken into account by calculating the stellar spectrum with \verb|PySME| at multiple $\mu$ angles equally spaced apart, following \cite{Yan2017}. Limb darkening is directly computed from model atmospheres using \verb|PySME| with a number of evenly spaced $\mu$ angles specified by the user input.

Additional input parameters of \verb|StarRotator| describe the properties of the transiting planet:
   \[
      \begin{array}{lp{0.8\linewidth}}
         a/R_\star  & scaled semi-major axis of orbit     \\
         e               & eccentricity                    \\
         \omega             & angle between the ascending node and periastron in deg             \\
         i_p & orbital inclination in deg \\
         \lambda          & projected orbital obliquity in deg     \\
         R_p/R_\star             & scaled planetary radius \\
         P         & period in days             \\
         T_c & mid-transit time in BJDTT - 2450000 \\

      \end{array}
   \]
\verb|StarRotator| calculates the stellar spectrum behind the transiting exoplanet and returns the spectral time-series over the entire transit event \citep[see][for relevant calculations]{Prinoth2024}. The combination of \verb|PySME| with \verb|StarRotator| allows a full transit time-series observation to be simulated, including distortions in the line profile caused by the exoplanet transit. We add noise to this simulated time-series observation corresponding to real HARPS observations in \cite{Prinoth2023} (see Section \ref{sec:w189}) and use it to test the method of isolating the narrow stellar spectrum, $F_{\star, N}$, as described below (see Section \ref{sec:method}).\\

\subsection{Isolating the stellar spectrum}
\label{sec:method}

As the planet transits across its host star, it obscures part of the starlight. We define the ratio between the in-transit and out-of-transit stellar spectrum as the residual spectrum, $R=R(t,\lambda)$. This is observed during a time-series observation with exposures taken at several times $t$:
\begin{equation}
\label{eq:res}
    R = \frac{F_\star-F_p}{F_\star} = \frac{F_{\text{it}}(t)}{F_{\text{oot}}},
\end{equation}
where $F_\star$ is the spectrum of the unobscured, disc-integrated star exhibiting rotation broadening in units of flux density (e.g. W/m$^2$). $F_p$ is the obscured local stellar spectrum\footnote{$F_p$ is not to be confused with the transmission spectrum of the atmosphere of the planet, which is not considered in this study.}, in the same units as $F_\star$, from the area behind the transiting planet at time $t$. In terms of observations, $F_{\text{it}}$ are the in-transit spectra, and $F_{\text{oot}}$ is the mean out-of-transit spectrum. \\
\\
\noindent First, we considered a planet transiting across a non-rotating star, with a non-broadened stellar spectrum and without centre-to-limb variation. The locally emitted stellar spectrum has the same line profile as the total stellar spectrum, $F_\star$, scaled by the transit depth, $D$. The residual spectrum for this non-rotating star is
\begin{equation}
\label{eq:nonrotating}
    R = \frac{F_\star - DF_\star}{F_\star} = 1-D.
\end{equation}

$D$ is equivalent to the relative transit depth in the light curve \citep{2013A&A...549A...9C}, which is dependent on the planetary radius $R_p/R_\star$ and the limb darkening parameters $u_1$ and $u_2$, based on the quadratic limb darkening model \citep{Kipping_2013}:
\begin{equation}
\label{eq:transit-depth}
    D = \frac{\Delta F}{F} = \left(\frac{R_p}{R_\star}\right)^2 \left(\frac{1-u_1(1-\mu)-u_2(1-\mu)^2}{1-u_1/3-u_2/6}\right).
\end{equation}
We chose to model the limb darkening with the quadratic law because it accurately reproduces the synthetically produced transit light curve from \verb|PySME| input spectra (see Section \ref{sec:w189}). This limb darkening model describes the intensity as
\begin{equation}
\label{eq:limbdarkening}
    \frac{I(\mu)}{I(1)} = 1-u_1(1-\mu)-u_2(1-\mu)^2,
\end{equation}
where $I(1)$ is the intensity at the centre of the stellar disc, which is normalised to 1. This varies as a function of position on the stellar disc $\mu(t)$, which is defined as
\begin{equation}
\label{eq:mu}
    \mu(t) = \cos \theta(t) = \sqrt{1-r(t)^2},
\end{equation}
where $\theta(t)$ is the centre-to-limb angle, and $r(t)$ is the radial distance from the centre of the star to the centre of the transiting planet, normalised to the stellar radius. From Equation~\eqref{eq:nonrotating}, it is clear that in the approximation of a non-rotating star without centre-to-limb variation, and fixed limb-darkening parameters, a transiting planet causes a decrease in flux that is independent of wavelength.\\

\begin{figure}
    \centering
    \includegraphics[width=0.5\textwidth]{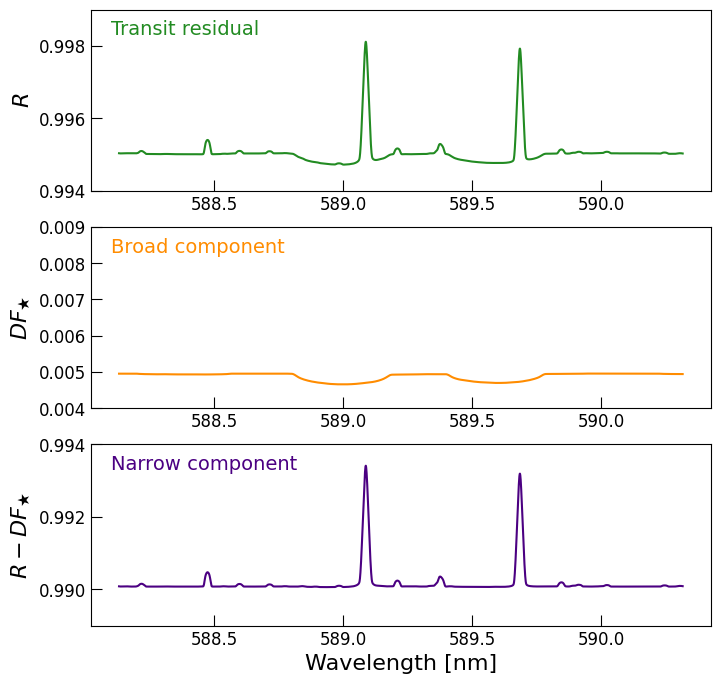}
    \caption{Decomposition of the residual spectrum into broad and narrow components. The top panel is the transit residual, $R$, that can be calculated following Equation~\eqref{eq:res}. This is further split into the broad component, (middle orange line) and the narrow component, (bottom purple line). The broadened component is the out-of-transit stellar spectrum scaled by the transit depth, $DF_\star$. The narrow component is the difference between the transit residual spectrum and the broadened component.}
    \label{fig:decomposed}
\end{figure}

\noindent We now turn to the case of a rotating star, for which line shape distortions appear in the in-transit spectrum, causing narrow and broad line-components to appear in the residual $R$. A simulation of such a residual $R$ is shown in Fig.~\ref{fig:decomposed} at the wavelengths of the Na doublet for an 8000 K A-type star, consistent with WASP-189 (see Section~\ref{sec:w189}). From visual inspection of the residuals in Fig.~\ref{fig:decomposed}, it is clear that there are two components: the narrow positive line core \citep[in some literature this is referred to as the Doppler shadow, e.g.][]{Gaudi2017,Hoeijmakers2020}, as well as the broader absorption line. In studies of transmission spectra of ultra-hot Jupiter atmospheres, both components have been approximated by fitting a composite Gaussian model to this line-shape, typically after application of cross-correlation \citep[e.g.][]{Cegla_2016,CasasayasBarris2018,Bourrier2020, Hoeijmakers2020,Prinoth2022}. In this paper, our objective is to use this signal $R$ for analysis of the stellar photosphere.

\begin{figure*}
    \centering
    \makebox[\textwidth][c]{\includegraphics[width=1.05\textwidth]{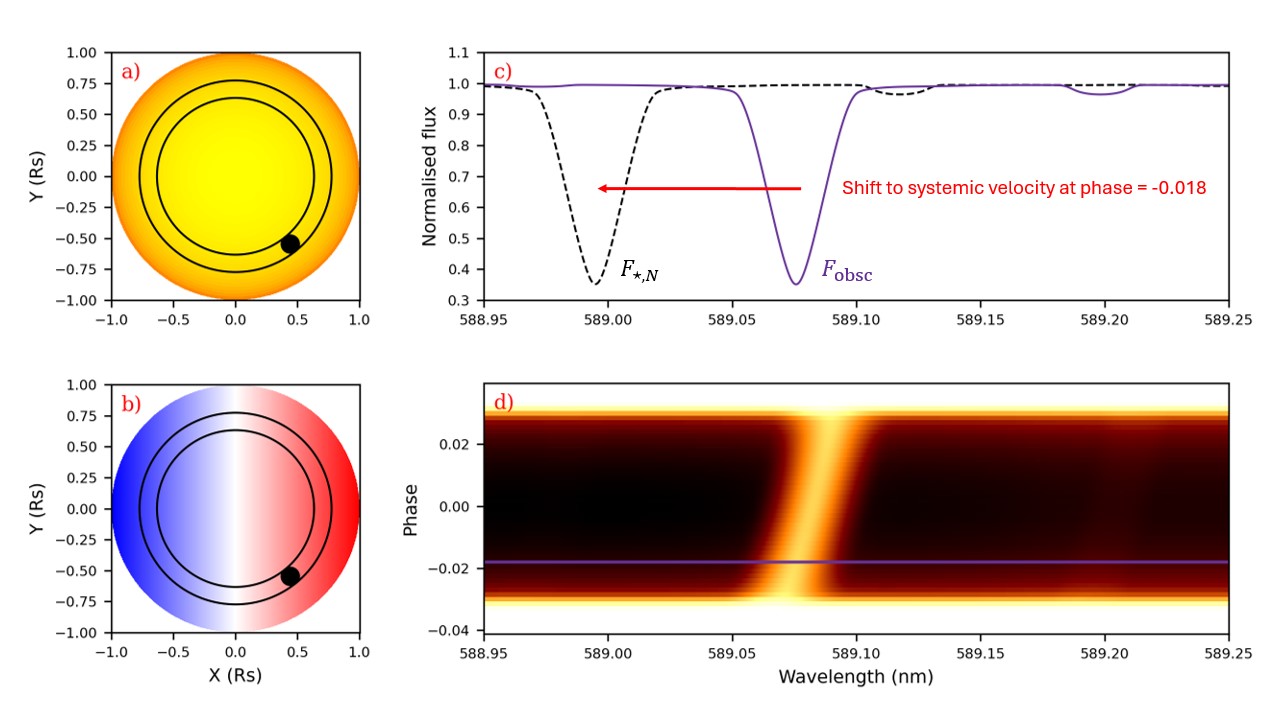}}
    \caption{Model isolated spectrum from a single in-transit exposure of WASP-189 b at phase $\phi=-0.018$. The spectrum has been cropped to show a single line of the Na doublet. Subfigures a) and b) show the position of the transiting planet at the current phase as a solid black circle over the star. The annulus of similar stellar composition is also indicated in these figures with solid black lines. The colour of the star in a) represents limb darkening, and b) represents the rotation of the star, with the respective blue and red Doppler shifts. Subfigure c) shows the momentary local obscured spectrum, $F_\text{obsc}$, in purple and the isolated spectrum, $F_{\star,N}$, with the black dashed line. d) contains the time-series in-transit residuals, $R$, with the horizontal purple line indicating the current planetary orbital phase. The colours represent the flux level, with greater flux towards the yellow spectrum, and less flux darker. Note that this residual spectrum is different from the extracted non-broadened spectrum above.}
    \label{fig:methodfig}
\end{figure*}
We noted that the broad component and the narrow component are equivalent to the scaled broadened out-of-transit spectrum and the non-broadened locally emitted spectrum respectively. This allowed the entire non-rotation broadened stellar spectrum to be isolated and used in spectral synthesis modelling, as demonstrated as follows.\\
\\
\noindent To isolate the narrow component, we followed a similar argument as for the non-rotating star. Now, $F_p$ does not have the same shape as $F_\star$, but is analogous to the narrow component as seen in Fig.~\ref{fig:decomposed}. The residual for a rotating star is
\begin{equation}
\label{eq:res-expanded}
    R = \frac{F_\star-DF_{\text{obsc}}}{F_\star}.
\end{equation}
From this, we extracted the narrow spectrum as
\begin{equation}
\label{eq:narrow}
    F_{\text{obsc}} = \frac{1-R}{D} F_{\star}.
\end{equation}
The thus obtained narrow spectrum $F_{\text{obsc}}$ is equal to the spectrum that is emitted locally behind the planet at any time during the transit event.

Each obscured spectrum $F_{\text{obsc}}$ is Doppler shifted according to the radial velocity of the patch of stellar disc situated directly behind the exoplanet, as seen in Figs.~\ref{fig:methodfig}b) and c). This was corrected for by applying the inverse Doppler shift:
\begin{equation}
\label{eq:doppler}
    \lambda_0 = \lambda \sqrt{\frac{1-\beta}{1+\beta}},
\end{equation}
where $\lambda_0$ is the wavelength axis in the rest frame, $\lambda$ is the observed wavelength axis, and $\beta$ is the Doppler factor, $v/c$. $v$ is the projected velocity directly behind the centre of the planet, assuming that the projected stellar rotation axis is aligned with the $y$-direction:
\begin{equation}
    v = x_p v_{\text{eq}}\sin i_\star,
\end{equation}
$x_p$ is the position of the planet's centre projected on the stellar disc, normalised to stellar radius. This shifts all local stellar spectra $F_{\obsc}$ into their rest-frames.\\
\\
Now consider that due to limb darkening and centre-to-limb variation, the local stellar spectrum varies with $\mu$, which is related to the radial distance, $r$, according to Equation~\eqref{eq:mu}. To reconstruct the total non-broadened spectrum of the star, $F_{\obsc}$ was integrated over the entire stellar disc. Assuming that emission from the stellar disc is point-symmetric (i.e. the absence of gravity darkening\footnote{Gravity darkening manifests itself as a temperature gradient from the stellar equator to the pole, breaking the point-symmetry assumed in this formalism. For typical exoplanet hosts, this temperature gradient can be as high as 700 K \citep{Cauley2022}, but this range would only be probed for planets with zero impact parameter.}), the annular region defined by the width and position of the planet (see Fig.~\ref{fig:methodfig}a)) emits a spectrum $F_{\obsc}$ throughout. The disc-integrated spectrum then becomes, taking into account the area of the annulus $A(r)= \pi ((r+R_p)^2 - (r-R_p)^2) = 4 \pi r R_p$ and the light intensity $I(r)$, according to Equation~\eqref{eq:limbdarkening}:

\begin{equation}
\label{eq:av_spectrum}
    F_{\star,N}(\lambda) = \frac{\sum_{r=0}^{R_\star} F_{\obsc}(r,\lambda)A(r)I(r)}{\sum_{r=0}^{R_\star} A(r)I(r)}.
\end{equation}
In this paper, we call $F_{\star,N}$ the isolated spectrum.

The centre of the stellar disc is not sampled during transits when a planet has a non-zero impact parameter, and this is generally the case, including for WASP-189 b. For this reason, we choose to approximate the spectrum of the central region as the spectrum of the inner-most sampled annulus. As typical limb darkening profiles are steeper towards the disc edges, we consider this to be an appropriate approximation for all systems where the impact parameter is not large enough to significantly affect the shape of the transit light curve.

The resulting quantity $F_{\star,N}$ approximates the stellar spectrum in the absence of rotation broadening. However, due to the non-zero extent of the obscuring planet, the area behind the planet still covers a small range of rotational velocities which leads to some broadening, approximately equal to $\frac{R_p}{R_{\star}} v_{\text{eq}}\sin i_\star $, which is much smaller than the rotational broadening in the out of transit spectrum. E.g. for WASP-189, $F_{\star,N}$ is broadened by 6-7 km/s. In real applications, a small amount of additional broadening may appear, due to micro/macro-turbulence \citep[2.7 km/s for WASP-189,][]{2020A&A...643A..94L} and finite instrumental spectral resolving power (2.7 km/s for HARPS) that we have ignored in this derivation.

\subsection{Flux calibration and continuum normalisation}
\label{sec:flux-calibration}
The theory described above assumes that the spectral time-series is flux calibrated. In reality, ground-based observations may be arbitrary scaled and vary in flux, so in real analyses, a normalisation is typically applied to $F(t)$. To ensure that Equation~\eqref{eq:res-expanded} holds in such cases, we carry out a rescaling of $F(t)$ by dividing out the mean of the spectrum, $\bar{F}(t)$, to obtain the normalised spectrum; and then reintroducing the broadband limb darkened light curve into the normalised time-series. This follows established practice when extracting obscured stellar line residuals \cite[see e.g.][]{Cegla_2016}. The observed residual $R'$ will be in the form
\begin{equation}
\label{eq:continuum}
    R' = \frac{(1-D) F(t)}{F_{\text{oot}}} \frac{\bar{F}_{\text{oot}}}{\bar{F}(t)}.
\end{equation}
In the theoretical case of flux-calibrated spectra, $\frac{\bar{F}(t)}{\bar{F}_{\text{oot}}}$ is trivially equal to $1-D$, while in the case of real observations, the spectra $F$ are scaled by an arbitrary continuum function $\alpha(\lambda)$ (e.g. because of instrumental throughput). Here that is divided out and disappears.\\

\noindent Additionally, spectral synthesis methods require the characterisation of a spectral continuum to reliably fit photospheric lines. Using our method, continuum normalisation can be applied to the final isolated spectrum, $F_{\star,N}$. This bypasses the problem of continuum normalisation with a rotation broadened spectrum, where severe line blending makes selecting continuum points for normalisation practically impossible (see Fig. \ref{fig:sim-broad-fit}). The term $\frac{1-R}{D}$ in Eq. \ref{eq:narrow} is approximately unity in the continuum. Therefore, by multiplying with the broadened out-of-transit stellar spectrum $F_\star$, the continuum of the obscured spectrum is scaled to the continuum level of $F_\star$ -- even if this continuum is not observed due to line blends (see Fig. \ref{fig:decomposed}).\\

\noindent An additional refinement of the continuum normalisation may need to be applied to $F_{\obsc}$ by dividing by the median to ensure the weighted average (Equation~\eqref{eq:av_spectrum}) is unbiased, but because the continuum of $\frac{1-R}{D}$ is close to 1, any errors introduced in this way are very small. We may thus proceed to apply typical continuum normalisation methods to $F_{\star,N}$ instead of $F_\star$.\\

\subsection{Fitting}
\label{sec:fitting}
We used the inbuilt fitting function in \verb|PySME| to obtain best-fit photospheric parameters using the isolated stellar spectrum, $F_{\star,N}$, as extracted above using the linelist extracted from VALD (see Section \ref{sec:method}). The solver minimises the least-square error between the input spectrum and the spectrum as modelled by \verb|PySME| for specified free parameters, $T_\text{eff}, \log g$, [Fe/H] and individual chemical abundances. The fitting process uses the SciPy \verb|dogbox| algorithm (\cite{voglis2004rectangular}; \cite{scipy}), which uses minimal function evaluations for convergence \citep{pySME}. This fitting algorithm was applied to the mean out-of-transit spectrum \citep[consistent with methods in][]{Prinoth2022}, as well as the derived $F_{\star,N}$, and also a synthetic spectrum with no rotation broadening. As well as the photospheric parameters, we were interested in fitting the $\alpha$ element abundances of Mg, Ca and Ti. The $\alpha$ element abundances are a metric often used in Galactic chemical evolution models to classify stellar populations \citep{matteucci1990}, and thus an interesting measure to provide beyond determining the stellar parameters. Furthermore, for future retrieval work on WASP-189\,b there is a need to know the stellar abundances for comparison with retrieved planetary abundances. \\
\\
The fitting process was as follows:
\begin{enumerate}
    \item Before fitting the photospheric parameters, we performed a sensitivity analysis to determine which lines in the spectrum are more sensitive to variations of global parameters, $T_{\text{eff}}$, $\log g$, [Fe/H] and $v\sin i_\star$, and, separately, individual elemental abundances. The latter was achieved by fixing the global parameters and synthesising new spectra with chemical abundances varying by $\pm 0.2$ dex. 

    \item Segments of approximately 0.5 -- 3 nm containing the most sensitive lines were selected for both the global parameters as well as for each elemental abundance. Regions where sensitive lines were strongly blended were discarded. In this way, 26 segments were selected to fit the global parameters, while individual species were fit using varying number of segments depending on the location of their strongest lines.

    \item For each selected segment, masks were created by manually selecting the wavelength regions containing the target lines (see Fig.~\ref{fig:sim-narrow-fit}). In this way, 66 masks over the entire spectrum containing 191 Fe lines (see Table~\ref{tab:fe-line-list}) were selected to fit the global parameters. Separate continuum masks were also created to allow \verb|PySME| to fit the arbitrary continuum level as described in Section ~\ref{sec:flux-calibration}.

    \item We conducted a fit to determine the optimal values of the global parameters, $T_{\text{eff}}$, $\log g$, [Fe/H] and $v\sin i_\star$ from the extracted $F_{\star,N}$ as well as their uncertainties.

    \item With global parameters $T_{\text{eff}}$, $\log g$ and $v\sin i_\star$ fixed to their best fit values, the selected lines were used to fit individual abundances separately.

    \item To estimate the systematic errors on individual abundances introduced by fixing the global parameters, we vary $T_\text{eff}$ and $\log g$ by their uncertainties (see point 3). This yields noise terms $\sigma_T$ and $\sigma_g$ that we add in squares to the statistical uncertainty, $\sigma_n$ reported by \verb|PySME|: $\sigma_T^2+ \sigma_g^2 + \sigma_n^2$. Note that $\sigma_n$ is usually small compared to the other terms.
\end{enumerate}

\section{WASP-189: Proof-of-concept}
\label{sec:w189}
Throughout this paper, we have used, as an example, the A-type star WASP-189 and its ultra-hot Jupiter WASP-189 b. The exoplanet was discovered by \citet{Anderson2018} with the WASP-South survey and TRAPPIST-South telescope, and confirmed with radial-velocity and Doppler tomography observations using HARPS and CORALIE spectrographs. This star is an ideal candidate to study with our method as it is a very bright ($M_V$ = 6.6 mag) fast rotator with a high projected rotational velocity $v\sin i_\star = 93.1$ km/s \citep{2020A&A...643A..94L}. Since its discovery in 2018, transits have been observed repeatedly with spectrographs \citep[e.g.][]{Prinoth2023}. As a proof-of-concept, we created a synthetic model of these observed time-series spectral observations using the method as described in Section ~\ref{sec:spectralsynthesis}.

The signal-to-noise (S/N) and wavelength range for the synthetically generated spectra were designed to match WASP-189 b transit spectra observed with the HARPS spectrograph \citep{HARPS}. HARPS has a S/N of 110 at 550 nm for a $M_V=6$ star observed for 1 minute and covers a wavelength range of 378 -- 691 nm over 72 spectral orders with a spectral resolution of 115 000. To ensure the calculation of the weighted average following Equation~\eqref{eq:av_spectrum} matches real transit spectra, the phases of each observation were taken to be the same as each night of HARPS observations in \cite{Prinoth2023}.\\
\\
\begin{table}
\centering
  \cprotect\caption[]{Input WASP-189 system parameters used to generate the synthetic spectrum with \verb|StarRotator|. All input parameters were taken from \citet{2020A&A...643A..94L}.}
     \label{param}
$ 
     \begin{array}{lc}
        \hline
        \hline
        \noalign{\smallskip}
        \text{Stellar Parameters}  \\
        \noalign{\smallskip}
        \hline
        \noalign{\smallskip}
        v_{\text{eq}} \text{ [km/s]}& 96.163   \\
        i_{\star} \text{ [deg]} & 75.5\\
        T_{\text{eff}} \text{ [K]}& 8000 \\
        \text{[Fe/H]} \text{ [dex]}& 0.29 \\
        \log g \text{ [log(cgs)]}& 3.9 \\
        \text{Number of $\mu$ angles} & 10\\
        \hline
        \hline
        \noalign{\smallskip}
        \text{Planetary Parameters}\\
        \noalign{\smallskip}
        \hline
        \noalign{\smallskip}
        a/R_{\star} & 4.60      \\
        e & 0 \\
        \omega \text{ [deg]} & 0 \\
        i_p \text{ [deg]} & 84.03\\
        \lambda \text{ [deg]} & 86.40\\
        R_p/R_{\star} & 0.07045  \\
        P \text{ [d]} & 2.724033 \\
        T_c \text{ [BJDTT - 2450000]} & 8926.5416960 \\
        \hline
     \end{array}
$
\end{table}

\noindent Following the method described in Section~\ref{sec:spectralsynthesis}, the stellar and planetary parameters from Table \ref{param} were used as inputs for \verb|StarRotator| and \verb|PySME| to interpolate spectra from a given model atmosphere and henceforth simulate the time-series synthetic spectra as observed with HARPS at the 3 epochs as stated in \cite{Prinoth2023}. The S/N values were estimated for each epoch from real time-series spectral observations in wavelength and time. Random Gaussian noise with a standard deviation corresponding to this S/N were added to each synthetic spectrum to mimic a real exposure.\\
\\
We required the radial velocity of the stellar disc directly behind the transiting planet to shift the spectra to the rest frame (see Equation~\eqref{eq:doppler}). \verb|StarRotator| computes the radial velocities over the stellar disc using a grid-based approach. The obscured spectra are shifted by the mean velocity of the grid cells blocked by the exoplanet.\\
\\
The transit light curve from StarRotator is calculated based on radiative transfer equations within PySME and not any specific limb darkening law. However, to calculate the transit depth, $D$, with Equation~\eqref{eq:transit-depth}, we require the limb darkening coefficients $u_1$ and $u_2$. We therefore obtained the limb darkening coefficients by fitting the quadratic limb darkening model to the light curve obtained using \verb|StarRotator|, using \verb|NumPyro| \citep{bingham2019pyro,phan2019composable} and \verb|Jax| \citep{jax2018github}. The fitted parameters are listed in Table~\ref{tab:lightcurve}, and are broadly consistent with the measured values from \cite{2022A&A...659A..74D}.\\
\\
\begin{table}
  \cprotect\caption[]{Fitted limb darkening coefficients to light curve generated with modelled spectra including centre-to-limb variation from \verb|PySME|.}
     \label{tab:lightcurve}
 $$ 
     \begin{array}{lcc}
        \hline
        \hline
        \noalign{\smallskip}
        \text{Coefficient} &\text{Fitted result}&\text{\cite{2022A&A...659A..74D}} \\
        \noalign{\smallskip}
        \hline
        \noalign{\smallskip}
        u_1  & 0.349 & 0.414^{+0.024}_{-0.022}\\
        \noalign{\smallskip}
        u_2& 0.256 & 0.155 ^{+0.032}_{-0.034}\\
        \noalign{\smallskip}
        \hline
     \end{array}
 $$ 
\end{table}

\begin{figure*}
\centering
\includegraphics[width=\textwidth]{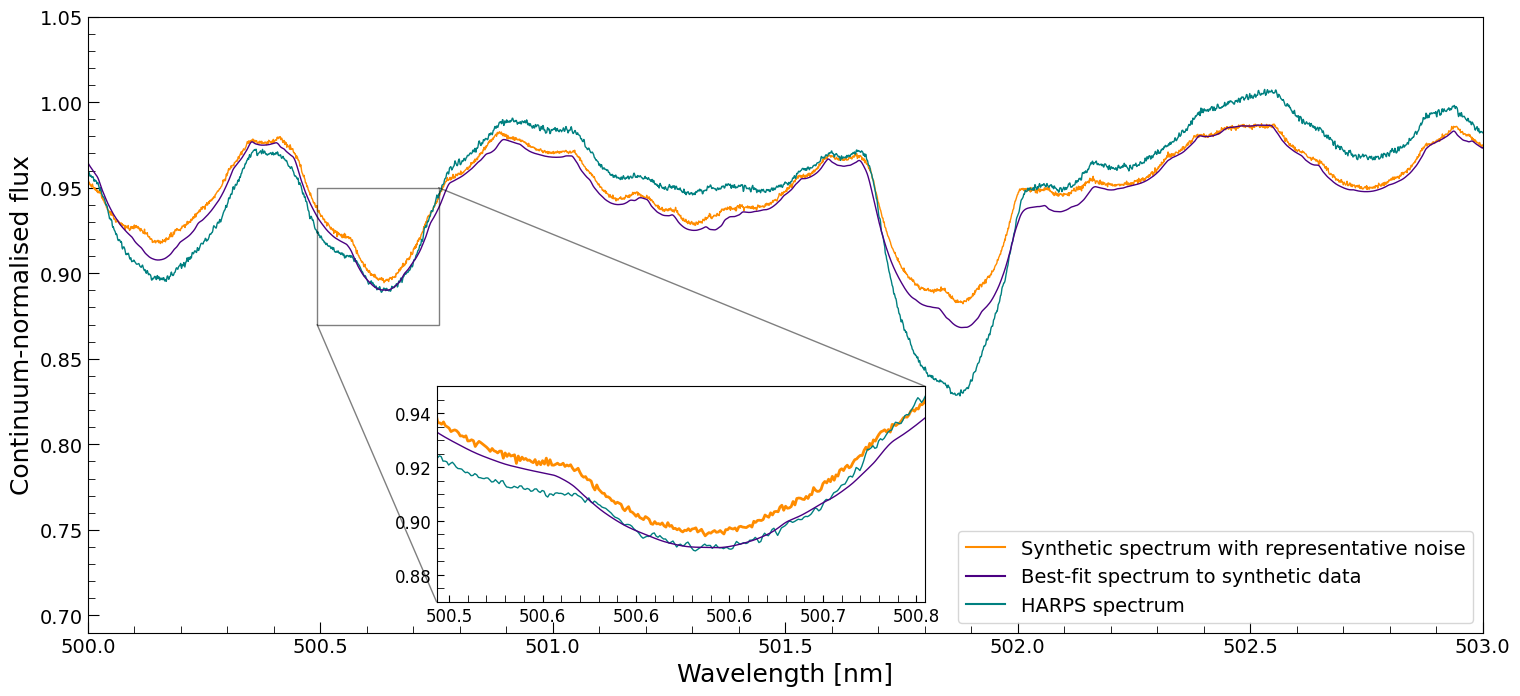}

\caption{Example segment of the mean out-of-transit stellar spectrum of WASP-189 from 3 nights of simulated HARPS observations (orange) with its best fit spectrum (purple). The real HARPS spectrum (green) has been included to show how the synthetically generated spectrum compares.}%
\label{fig:sim-broad-fit}
\end{figure*}

\begin{table*}
  \cprotect\caption[]{Results from the fitted model of synthetic spectra of WASP-189 compared with the input parameters. The fit was carried out on the mean out-of-transit broadened spectrum from 3 transits, the isolated spectrum $F_{\star,N}$, calculated from 3 transits, the true isolated spectrum, $F_{\star,N}$ with arbitrary noise with S/N=50 added and a non-broadened spectrum with the same arbitrary noise. Individual chemical abundances are reported relative to solar abundances from \cite{Grevesse2007}. Note that the uncertainties are statistical uncertainties reported by \verb|PySME| and do not take into account any propagated errors into consideration.}
    \label{tab:sim-results}
    \centering
     \begin{tabular}{p{0.15\linewidth}ccccc}
        \hline
        \hline
        \noalign{\smallskip}
        Parameter      &\text{Input}&  \text{Broadened} & $F_{\star,N}$ \text{ (3 transits)} & $F_{\star,N}$ \text{ (S/N = 50)} & Non-broadened (S/N=50)\\
        \noalign{\smallskip}
        \hline
        \noalign{\smallskip}
        $T_{\text{eff}}$ [K] & 8000 & $8400\pm30$  & $8100 \pm 50$ & $7990\pm10$&$8000.0\pm3.3$\\
        $\log g$ & 3.9& $4.05\pm0.02$ & $3.94 \pm 0.05$&$3.87\pm0.01$&$3.900\pm0.003$\\
        $v\sin i$ [km/s] & & $97.1\pm0.2$ & $6.51 \pm 0.07$&$6.83\pm0.02$&$0.00\pm0.07$\\
        \noalign{\smallskip}
        \hline
        \noalign{\smallskip}
        \text{[Fe/H] }[dex]&0.29 & $0.67\pm0.02$ & $0.42 \pm 0.03$ &$0.304\pm0.006$&$0.288\pm0.002$\\
        \text{[Mg/H]} & 0.29& $0.921\pm 0.008$&$0.44\pm 0.02$&$0.289\pm0.006$&$0.30\pm0.01$\\
        \text{[Ca/H]} &0.29& $1.234\pm0.005$&$0.43\pm0.02$&$0.304\pm0.005$&$0.283\pm0.008$\\
        \text{[Ti/H]} & 0.29& $1.79\pm0.01$&$0.37\pm0.03$&$0.263\pm0.009$&$0.298\pm0.01$\\
        
        \noalign{\smallskip}
        \hline
     \end{tabular}
 
\end{table*}
\noindent Figure~\ref{fig:sim-broad-fit} shows the mean out-of-transit broadened stellar spectrum of WASP-189, which would usually be used to fit photospheric parameters. We carried out a fit of the broadened out-of-transit spectrum using \verb|PySME| (see Section~\ref{sec:fitting}) and due to the high S/N, the measurements were precise. However the fitted values were systematically offset (see Table~\ref{tab:sim-results} and Fig. \ref{fig:sim-broad-fit}). This is because continuum points are difficult to identify within a wavelength segment, affecting continuum fitting and introducing significant errors when fitting lines of the significantly rotation broadened ($v\sin i_\star \approx 100$ km/s) spectrum as all the lines are blended, skewing the results of the fitted spectrum -- even for a fully synthetically modelled spectrum.

\begin{figure*}
\centering
\includegraphics[width=\textwidth]{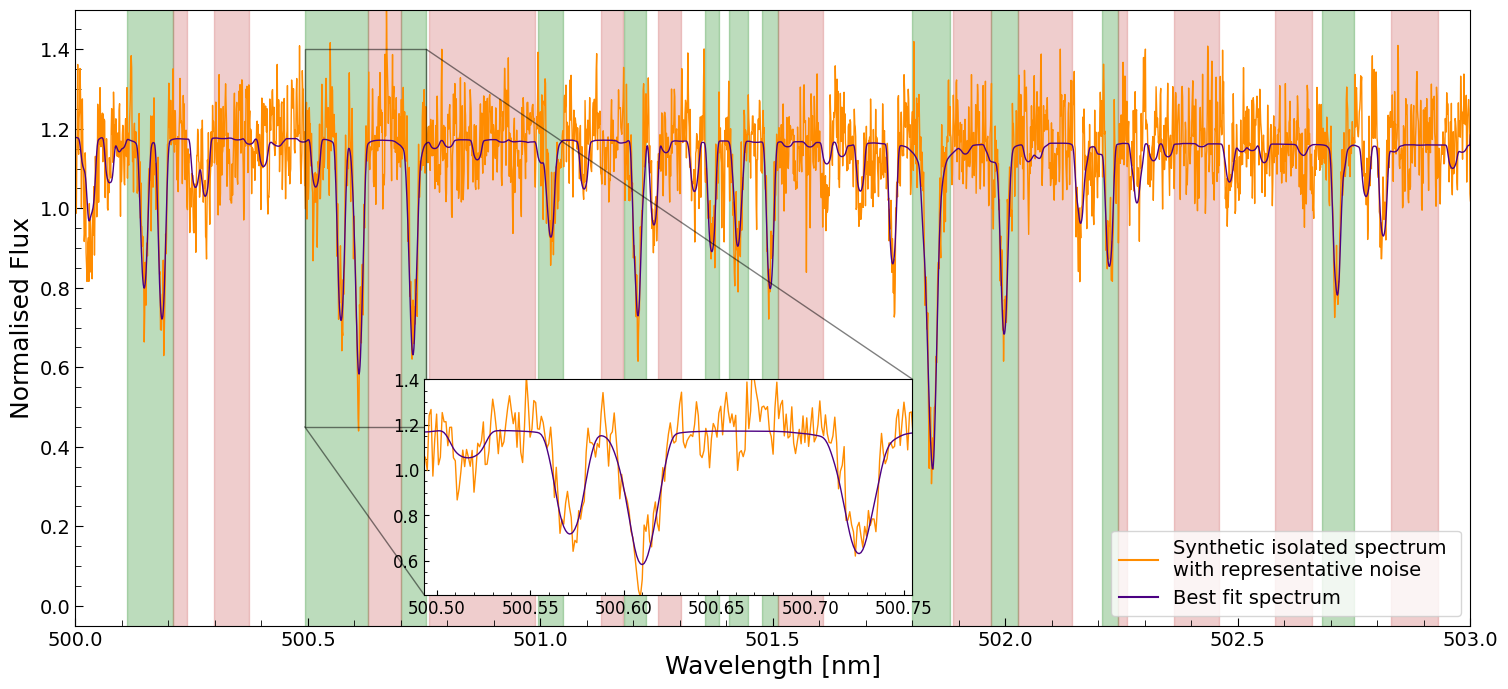}

\caption{Example segment of the average isolated stellar spectrum, $F_{\star,N}$, from 3 WASP-189 b transits from simulated HARPS observation (orange) with its best-fit spectrum (purple). The shaded green regions marks the pixels used for fitting various parameters and abundances. The red regions marks the pixels used for continuum fitting.}%
\label{fig:sim-narrow-fit}
\end{figure*}

On the other hand, the isolated spectrum, $F_{\star,N}$ shows clear spectral lines and a flat continuum level, however at a cost of decreased S/N (see Fig.~\ref{fig:sim-narrow-fit}), which is the main limitation of our method. From a single transit observation of WASP-189 b with HARPS, the S/N of $F_{\star,N}$ was approximately 5. This increases to approximately 10, for three stacked transits. When fitting the photospheric parameters, the statistical uncertainties were increased, however the results were more accurate.

\begin{figure}
\centering
\includegraphics[width=0.5\textwidth]{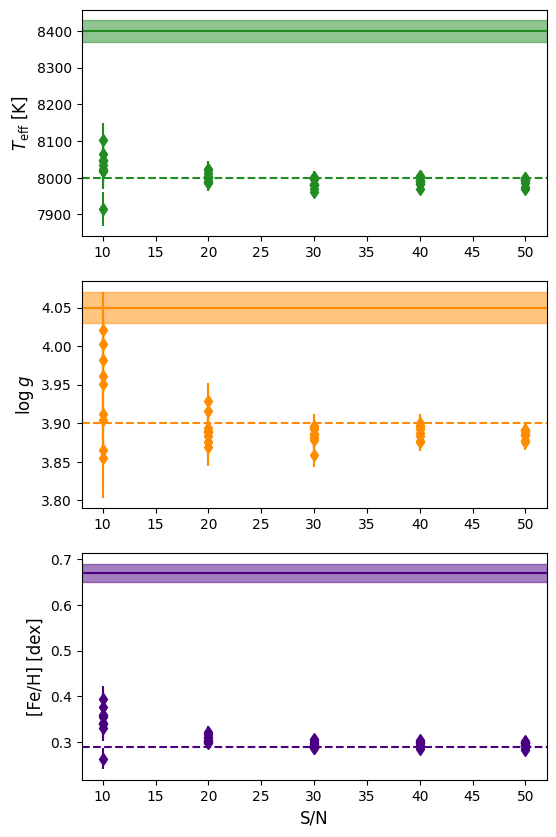}
   \caption{Fitted global photospheric parameters ($T_{\text{eff}}, \log g, $ [Fe/H], from top to bottom) of the true synthetic isolated spectrum, $F_{\star,N}$, with varying S/N. The dashed horizontal line is the true value of the corresponding parameter. The shaded horizontal region is the fitted result from the synthetic broadened fit.}
      \label{fig:resultsnr}
\end{figure}

To determine how well we can fit the isolated spectrum in an ideal scenario with decreased noise, we conducted a separate analysis where we varied the noise added to the synthetic isolated spectrum, $F_{\star,N}$. We added random Gaussian noise with S/N of 10, 20, 30, 40 and 50 to obtain datasets of the same spectrum with varying noise levels. This was repeated 10 times to produce a spread of datasets with different realisations of random Gaussian noise, but with the same S/N. Following the same fitting procedure as in Section \ref{sec:fitting}, the results for the global parameters $T_\text{eff}, \log g$ and [Fe/H] are presented in Fig.~\ref{fig:resultsnr}, including the spread from fitting each of the 10 dataset with the same S/N. The best-fit parameters were observed to converge to the true value at higher S/N $\approx 40$ (see Fig.~\ref{fig:resultsnr}). This indicates that with a sufficiently high S/N, it is possible to use the isolated spectrum, $F_{\star,N}$, to fit photospheric parameters with increased accuracy and reliability compared to the out-of-transit stellar spectrum with a rotation broadening of $v_{\text{eq}} = 96.163$ km/s.

\section{Results}
\begin{figure*}
\centering
\includegraphics[width=\textwidth]{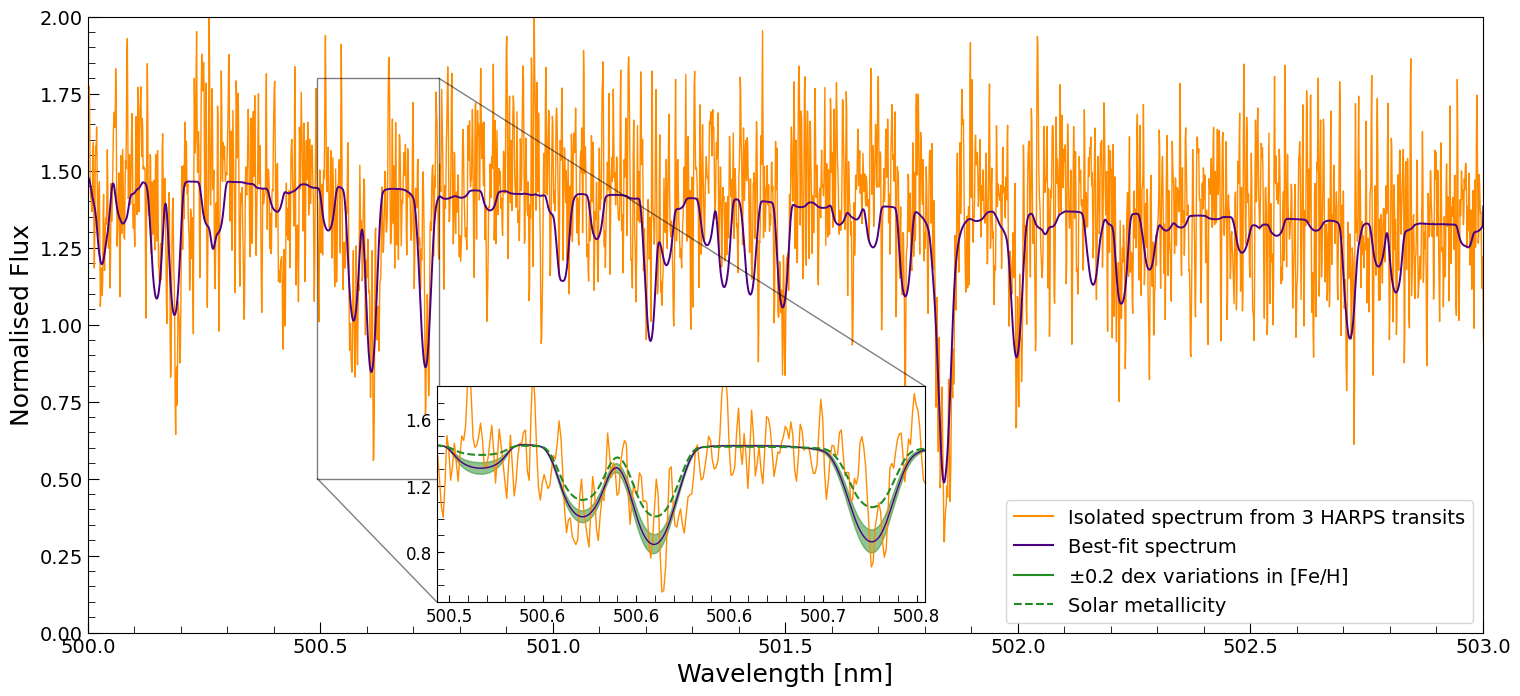}

\caption{Average isolated spectrum, $F_{\star,N}$, calculated from 3 real HARPS transits (orange) and its corresponding best-fit spectrum (purple) with $T_\text{eff}$ and $\log g$ fixed to parameters from \cite{2020A&A...643A..94L}. The shaded green region in the inset indicates $\pm0.2$ variations in metallicity, [Fe/H]. The dotted green line indicates the spectrum with solar metallicity, [Fe/H] = 0.}%
\label{fig:harps-fit}
\end{figure*}

Best-fit parameters to the stellar spectra, $F_{\star}$, (broadened and non-broadened), as well as the isolated spectra, $F_{\star,N}$, derived from the synthetic data are shown in Table \ref{tab:sim-results} (see also Section \ref{sec:fitting}). Fits to broadened stellar lines converge to systematically deviant values, but fits to the isolated spectrum can more accurately fit these parameters. The reported uncertainties are solely based on the least-squares algorithm used by \verb|PySME| to fit the spectra, and do not take into account correlations between parameters. We therefore carried out these fits several times, for varying S/N ratios, to demonstrate that statistically, the best-fit parameters converge to the true input values of the generated model (see Fig. \ref{fig:resultsnr}).\\
\\
In addition to testing our method with synthetic models, we have also isolated the narrow stellar spectrum of WASP-189 from real HARPS observations previously published in \cite{Prinoth2023}, see Fig. \ref{fig:harps-fit}. We obtained best-fit parameters shown in Table \ref{tab:HARPS-results}, with the best-fit spectrum overplotted in Fig. \ref{fig:harps-fit}. Two separate abundance fits were performed: first with fixing $T_{\text{eff}}$ and $\log g$ to the fitted parameters from a global fit on the same isolated spectrum, and another with fixing $T_{\text{eff}}$ and $\log g$ to literature values reported by \cite{2020A&A...643A..94L}. Fixing $T_{\text{eff}}$ and $\log g$ to independently measured values is expected to be the strongest use-case for the method we present here: Such parameters will generally be available from other photometric or spectroscopic measurements. Given these, accurate measurements of the abundances of individual elements can be obtained more reliably than from the rotation-broadened spectrum. This is because the isolated spectrum, $F_{\star,N}$ exhibits significantly less lines blending and a more identifiable continuum level, allowing lines of target elements to be fit more accurately.

When fixing $T_\text{eff}$ and $\log g$ to the values measured by \cite{2020A&A...643A..94L}, we find a metallicity of $0.50\pm0.05$ dex. This means that our measurement indicates that WASP-189 is a metal-rich star, with a higher metallicity than previously reported \cite[$0.29\pm0.13$ dex][]{2020A&A...643A..94L}. Fig. \ref{fig:harps-fit} shows a comparison between our measured metallicity of [Fe/H] = 0.5 dex with model spectra varying [Fe/H] by $\pm0.2$ dex. This shows that despite the relatively high noise level, \verb|PySME| is able to converge robustly. This measured metallicity is consistent with expectations that metal-rich stars are more likely to host gas giants \citep{Fischer2005}.\\
\\
Not only do we measure a new value for the metallicity of WASP-189, but we are also able to fit abundances of other elements. As a proof of concept, we have focused on measuring the abundances of $\alpha$ elements Mg, Ca and Ti. We report the abundances of these elements in Table \ref{tab:HARPS-results}. The individual chemical abundances for $\alpha$ elements we report are also consistent with previous surveys reporting that high metallicity stars have less $\alpha$ enhancement \citep{Buder2021}.\\

Previous studies struggle to reliably determine abundances of fast-rotating stars due to extreme line blending making it difficult to identify lines of target species \citep[e.g.][]{Ansari1992,Takeda2009}. This difficulty applies to all methods used to fit abundances, including spectral line fitting and equivalent width calculations. \cite{Takeda2009} states that the reported abundances for the fast rotators ($v\sin i_\star \gtrapprox 100$ km/s) in their study are unreliable due to the lack of unblended lines available for fitting. By isolating the stellar spectrum with the help of an exoplanet companion, we prove that many more lines can be accessed (see Fig.~\ref{fig:sim-narrow-fit} and Table~\ref{tab:fe-line-list}). A fitting method of the user's choice can then be applied to the isolated spectrum, $F_{\star,N}$.\\

When fitting $v \sin i_\star$ to real observational data, we report an increased value compared to the synthetic case. This additional broadening of $v\sin i_\star \approx 3$ km/s is likely due to other effects such as spectrograph line spread functions being absorbed into the total $v\sin i_\star$. The instrument's exposure time also introduces blurring from the planet moving across the stellar disc, broadening the spectrum by a few km/s.\\

Our method is limited by the S/N ratio of current observations. The isolated stellar spectrum, $F_{\star,N}$, has a significantly lower S/N than the mean out-of-transit broadened spectrum (see Fig. \ref{fig:sim-narrow-fit}). This limits how precisely we can measure stellar abundances. High-resolution spectrographs on bigger telescopes (e.g. ANDES on the ELT) will enable observations of $F_{\star,N}$ at higher S/N ratios, leading to much smaller uncertainties on the photospheric parameters and elemental abundances (see Fig. \ref{fig:resultsnr}), making this method of isolating the stellar spectrum potentially very powerful in the future. This study was limited to three available HARPS observations, but more in-transit observations of the target star will also help to improve the final S/N. Thus, if provided with a few time-series spectra of sufficient S/N, it will be beneficial to apply our method of isolating the narrow stellar spectrum to measure individual chemical abundances.

\begin{table}
  \caption[]{Fitted results for WASP-189 from 3 stacked transits observed with HARPS. The first column of fitted abundance results assume fixed $T_{\text{eff}}$, $\log g$ and $v\sin i_\star$ from a separate global fit. The second column of fitted abundances assume fixed $T_{\text{eff}}$ and $\log g$ from \cite{2020A&A...643A..94L}. The individual abundances are reported relative to solar abundances from \cite{Asplund2009}.}
    \label{tab:HARPS-results}
    \centering
     \begin{tabular}{p{0.25\linewidth}cc}
        \hline
        \hline
        \noalign{\smallskip}
        Parameter      &Fitted result & Fix to \text{\citet{2020A&A...643A..94L}} \\
        \noalign{\smallskip}
        \hline
        \noalign{\smallskip}
        $T_{\text{eff}} $ [K]& $7900\pm 100$ &$8000$\\
        $\log g$ & $3.4\pm0.1$&$3.9$\\
        \noalign{\smallskip}
        \hline
        \noalign{\smallskip}
        $v\sin i_\star$ [km/s]& $9.9\pm0.1$ & $9.8\pm0.2$\\
        \noalign{\smallskip}
        \hline
        \hline
        \noalign{\smallskip}
        \multicolumn{3}{l}{Fitted abundance [dex]} \\
        \noalign{\smallskip}
        \hline
        \noalign{\smallskip}
        \text{[Fe/H]} & $0.5\pm0.1$ &$0.56\pm0.06$\\
        \text{[Mg/H]} &$0.65\pm0.07$&$0.53\pm0.07$\\
        \text{[Ca/H]} &$0.45\pm0.07$&$0.44\pm0.07$\\
        \text{[Ti/H]} &$0.32\pm0.09$&$0.48\pm0.09$\\
        \hline
     \end{tabular}
 
\end{table}

\section{Conclusions}
Line blending in rotation broadened stellar spectra poses a problem when trying to measure the photospheric properties and individual chemical abundances in fast-rotating A-type stars. In this work, we have presented a new method to isolate the non-broadened stellar spectrum from the flux obscured by a transiting exoplanet, commonly referred to as the Doppler shadow. This spectrum, $F_{\star,N}$, is isolated from transit spectral time-series observations, and allows fitting of photospheric parameters using standard spectral synthesis methods. We demonstrated our proof-of-concept using synthetic spectra of WASP-189 generated with \verb|PySME| and \verb|StarRotator|. We have shown that even at the expense of decreased S/N, the narrow isolated spectrum, $F_{\star,N}$, has a much more identifiable continuum level and less severe line blending, allowing us to more precisely measure stellar abundances.\\

We have applied this method to real observational data from the combination of three transits observed by the HARPS spectrograph, measuring higher metal abundances than previously reported by \cite{2020A&A...643A..94L}. With instruments with improved S/N, such as the MAROON-X spectrograph on Gemini North or larger telescopes, or by using spectra from additional transit events, it will be possible to more tightly constrain these values.\\

These higher precision stellar abundances have important implications for the study of ultra-hot Jupiters, which tend to orbit young fast-rotating A-type host stars. Knowing the composition of the host star more precisely will allow for more reliable transmission spectroscopy for hot Jupiters, that is ultimately used to understand formation and evolution mechanisms involved in these systems, based on how the composition of the planet compares to the composition of the host star. This method can be applied to other hot Jupiter systems, such as KELT-9, MASCARA-2 or MASCARA-4, or to other systems involving fast-rotating stars and a companion that partially obscures the stellar disc during a transit event.\\ 
 
\begin{acknowledgements}
M.L. acknowledges financial support from the Lund Global Scholarship and the Eleanor Sophia Wood Postgraduate Research Travelling Scholarship. B.P. acknowledges financial support from The Fund of the Walter Gyllenberg Foundation. B.T. acknowledges the financial support from the Wenner-Gren Foundation (WGF2022-0041). This work is based in part on observations collected at the European Southern Observatory under ESO programme 107.22QF. This research has made use of the services of the ESO Science Archive Facility.
\end{acknowledgements}

%
  \bibliographystyle{aa} 
  \bibliography{literature} 

\begin{appendix} 
\section{Line selection}

{
\begin{center}
\topcaption{Line data for Fe}
\label{tab:fe-line-list}
\vspace{5mm}
\tablefirsthead{\hline
\hline
\noalign{\smallskip}
Wavelength in air [Å] & $E_\text{exc}$ [eV] & $\log gf$ \\
\noalign{\smallskip}
\hline
\noalign{\smallskip}}

\tablehead{\small\sl ...continued\\
\hline
\noalign{\smallskip}
Wavelength in air [Å] & $E_\text{exc}$ [eV] & $\log gf$ \\
\noalign{\smallskip}
\hline
\noalign{\smallskip}}
\begin{supertabular}{ccc}

Fe 1 & & \\
4654.49762 & 1.5574 & -2.783\\
4654.60500 & 3.6025 & -1.077\\
4654.63112 & 3.2112 & -1.559\\
4657.31848 & 2.4242 & -4.388\\
4657.38487 & 4.3865 & -2.762\\             
4666.60375 & 5.3448 & -0.181\\
4702.59763 & 4.3714 & -2.285\\
4702.66974 & 3.0469 & -4.960\\
4702.70049 & 4.6520 & -3.677\\
4702.91309 & 3.2367 & -2.830\\
4708.73490 & 4.4153 & -3.282\\
4714.35904 & 4.6070 & -1.691\\
4714.36837 & 5.3717 & -0.478\\
4783.36035 & 4.0758 & -4.433\\
4783.45969 & 5.0674 & -2.140\\
4786.80668 & 3.0173 & -1.606\\
4957.29779 & 2.8512 & -0.408\\
4957.59599 & 2.8083 & 0.233\\
4957.68179 & 4.1909 & -0.937\\
4978.35629 & 4.8349 & -2.195\\
4978.60272 & 3.9841 & -0.877\\
4978.69124 & 4.0758 & -1.009\\
4983.24993 & 4.1544 & -0.123\\
4983.32669 & 3.6416 & -4.491\\
4983.34110 & 4.6070 & -1.827\\
4983.85213 & 4.1034 & -0.005\\
4988.94928 & 4.1544 & -0.530\\
4988.95010 & 4.1544 & -0.890\\
4990.69267 & 4.9556 & -2.986\\
4990.90520 & 4.9556 & -3.902\\
4991.11052 & 2.7785 & -4.541\\
4991.14067 & 3.2367 & -4.915\\
4991.26751 & 4.1909 & -0.378\\
4991.26830 & 4.1909 & -0.670\\
4993.24894 & 2.8316 & -4.819\\
4993.48513 & 2.5881 & -5.923\\
5001.22995 & 4.6520 & -2.402\\
5001.42436 & 3.0713 & -4.694\\
5001.86303 & 3.8816 & 0.010\\
5001.86453 & 4.5485 & -1.242\\
5005.71197 & 3.8835 & 0.033\\
5005.86913 & 2.1759 & -7.173\\
5006.11078 & 4.8349 & -3.359\\
5006.11830 & 2.8325 & -0.638\\
5007.24664 & 3.9433 & -2.575\\
5007.27398 & 4.1034 & -0.204\\
5010.26883 & 4.5585 & -1.233\\
5010.29921 & 2.5592 & -5.104\\
5011.90780 & 4.9880 & -3.655\\
5011.96584 & 3.0176 & -5.442\\
5012.06761 & 0.8590 & -2.642\\
5012.15657 & 4.1909 & -0.977\\
5014.83927 & 5.0674 & -3.354\\
5014.94216 & 3.9433 & -0.303\\
5018.02679 & 3.6352 & -3.096\\
5018.35098 & 3.8816 & -3.133\\
5018.39960 & 5.1124 & -2.010\\
5018.41270 & 3.2367 & -3.173\\
5018.61072 & 4.6070 & -2.108\\
5018.79086 & 4.3013 & -3.300\\
5019.73083 & 3.9841 & -2.126\\
5022.16506 & 4.5585 & -2.963\\
5022.23520 & 3.9841 & -0.530\\
5022.36993 & 4.5931 & -3.035\\
5027.12225 & 4.1544 & -0.289\\
5027.22539 & 3.6398 & -1.885\\
5027.29238 & 3.6398 & -2.858\\
5027.34269 & 3.9841 & -3.154\\
5031.03752 & 3.6416 & -2.312\\
5031.03778 & 3.5465 & -5.741\\
5031.10462 & 3.6416 & -4.262\\
5040.85168 & 4.2562 & -0.606\\
5040.90608 & 4.2833 & -0.445\\
5041.07130 & 0.9582 & -3.087\\
5041.68501 & 4.8349 & -4.163\\
5041.72595 & 3.6946 & -3.887\\
5041.75544 & 1.4849 & -2.203\\
5041.84851 & 4.2833 & -0.857\\
5064.95193 & 4.2562 & -1.252\\
5065.01737 & 4.2562 & 0.005\\
5065.11437 & 4.5931 & -3.385\\
5065.16672 & 4.9880 & -4.162\\
5065.19239 & 3.6416 & -1.513\\
5068.76549 & 2.9398 & -1.042\\
5074.71968 & 3.5465 & -3.185\\
5074.74751 & 4.2204 & -0.200\\
5078.90557 & 3.9286 & -5.414\\
5078.91486 & 3.0176 & -5.695\\
5078.97395 & 4.3013 & -0.291\\
5079.17703 & 5.0095 & -3.292\\
5079.22219 & 2.1979 & -2.067\\
5080.35172 & 3.6946 & -2.190\\
5096.99717 & 4.2833 & -0.268\\
5098.47210 & 3.2515 & -4.694\\
5098.57221 & 3.9286 & -0.779\\
5098.66634 & 4.3714 & -3.911\\
5098.69807 & 2.1759 & -2.026\\
5125.11626 & 4.2204 & 0.002\\
5125.11710 & 4.2204 & -0.140\\
5125.25815 & 4.9556 & -2.895\\
5142.44512 & 4.5585 & -1.506\\
5142.49353 & 4.3013 & -0.739\\
5142.54009 & 4.2562 & -0.316\\
5142.85252 & 5.0095 & -2.884\\
5142.92819 & 0.9582 & -3.080\\
5146.30640 & 4.3714 & -2.030\\
5148.04193 & 4.2833 & -0.315\\
5148.19119 & 3.1973 & -5.660\\
5148.22830 & 4.2562 & -0.242\\
5150.83893 & 0.9901 & -3.003\\
5154.01733 & 4.9556 & -3.871\\
5154.10024 & 3.8816 & -1.712\\
5162.14483 & 5.0674 & -4.148\\
5162.27198 & 4.1777 & 0.020\\
5162.37595 & 2.6085 & -4.601\\
5165.40980 & 4.2204 & -0.384\\
5167.48785 & 1.4849 & -1.118\\
5168.84298 & 2.9904 & -5.433\\
5168.86436 & 3.3009 & -4.960\\
5168.89723 & 0.0516 & -3.969\\
5169.29841 & 4.0758 & -2.487\\
5169.33235 & 4.9130 & -3.608\\
5169.47108 & 4.2944 & -3.033\\
5171.46306 & 4.9850 & -4.165\\
5171.59602 & 1.4849 & -1.793\\
5171.67173 & 3.6416 & -1.933\\
5183.27011 & 4.8349 & -4.114\\
5183.46496 & 3.1110 & -4.937\\
5183.54478 & 5.0638 & -3.901\\
5183.69476 & 3.3014 & -6.679\\
5183.76975 & 5.8283 & -1.397\\
5184.14635 & 5.0331 & -2.685\\
5184.17619 & 4.3125 & -2.207\\
5184.25335 & 4.9913 & -1.830\\
5184.26572 & 4.2833 & -0.707\\
5184.26610 & 4.2833 &-1.000\\
5188.49163 & 4.4153 & -4.878\\
5194.94115 & 1.5574 & -2.090\\
5195.47139 & 4.2204 & -0.085\\
5202.25527 & 4.2562 & -0.472\\
5202.33540 & 2.1759 & -1.838\\
5226.40155 & 2.9488 & -5.181\\
5226.86117 & 3.0385 & -0.555\\
5226.89615 & 4.2944 & -1.906\\
5227.14950 & 2.4242 & -1.352\\
5227.18885 & 1.5574 & -1.228\\
5254.95507 & 0.1101 & -4.764\\
5254.97607 & 4.2944 & -1.742\\
5261.66985 & 4.4153 & -4.884\\
5262.24704 & 5.0995 & -3.060\\
5264.20971 & 5.1055 & -3.950\\
5264.37936 & 4.9130 & -2.022\\
5265.41858 & 4.3125 & -3.617\\
5265.72449 & 6.0383 & -0.953\\
5265.90977 & 4.7331 & -3.664\\
5265.93557 & 2.5592 & -3.854\\
5265.95692 & 3.6352 & -3.331\\
5270.05762 & 3.6352 & -2.643\\
5270.34629 & 5.6491 & -1.451\\
5270.35601 & 1.6079 & -1.339\\
5455.10307 & 3.2515 & -4.650\\
5455.38855 & 4.4733 & -0.747\\
5455.45344 & 4.3201 & 0.329\\
5455.60885 & 1.0111 & -2.091\\
5602.76727 & 4.1544 & -1.135\\
5602.80903 & 5.0674 & -1.645\\
5602.94468 & 3.4302 & -0.850\\
5657.66471 & 5.0638 & -2.637\\
5687.85661 & 4.2944 & -3.476\\
5914.11126 & 4.6076 & -0.375\\
5914.20048 & 4.6076 & -0.131\\
6137.46479 & 4.3013 & -3.793\\
6137.49795 & 3.3320 & -2.512\\
6137.69088 & 2.5881 & -1.403\\
6318.01746 & 2.4534 & -2.262\\
6399.99991 & 3.6025 & -0.290\\
Fe 2& & \\
4648.93295 & 2.5827 & -4.565\\
4656.97613 & 2.8910 & -3.610\\
4666.74949 & 2.8281 & -3.368\\
4993.35020 & 2.8067 & -3.640\\
5005.52199 & 0.3519 & -8.492\\
5018.43563 & 2.8910 & -1.220\\
5020.24431 & 0.3865 & -8.436\\
5150.92864 & 2.8555 & -4.483\\
5169.02818 & 2.8910 & -1.250\\
5171.63106 & 2.8067 & -4.805\\
5254.91999 & 3.2305 & -3.336\\
5261.63274 & 0.3013 & -7.658\\
5657.92439 & 3.4245 & -4.112\\
6238.38593 & 3.8887 & -2.754\\
6317.98871 & 5.5107 & -2.155\\

\end{supertabular}
\end{center}
}

\begin{center}
\topcaption{Line data for Ca}
\vspace{5mm}
\tablefirsthead{\hline
\hline
\noalign{\smallskip}
Wavelength in air [Å] & $E_\text{exc}$ [eV] & $\log gf$ \\
\noalign{\smallskip}
\hline
\noalign{\smallskip}}

\tablehead{\small\sl ...continued\\
\hline
\noalign{\smallskip}
Wavelength in air [Å] & $E_\text{exc}$ [eV] & $\log gf$ \\
\noalign{\smallskip}
\hline
\noalign{\smallskip}}
\begin{supertabular}{ccc}
Ca 1&&\\
5188.84400 & 2.9325 & -0.075\\
5260.38700 & 2.5213 & -1.719\\
5261.70400 & 2.5213 & -0.579\\
5264.23700 & 2.5230 & -0.574\\
5265.55600 & 2.5230 & -0.113\\
5349.46500 & 2.7090 & -0.310\\
5581.96500 & 2.5230 & -0.555\\
5588.74900 & 2.5257 & 0.358\\
5590.11400 & 2.5213 & -0.571\\
5594.46200 & 2.5230 & 0.097\\
5598.48000 & 2.5213 & -0.087\\
5601.27700 & 2.5257 & -0.523\\
5857.45100 & 2.9325 & 0.240\\
6102.43914 & 2.5230 & -3.188\\
6102.72300 & 1.8793 & -0.793\\
6122.21700 & 1.8858 & -0.316\\
6161.29700 & 2.5230 & -1.266\\
6162.17300 & 1.8989 & -0.090\\
6163.75500 & 2.5213 & -1.286\\
6166.43900 & 2.5213 & -1.142\\
6169.04200 & 2.5230 & -0.797\\
6169.56300 & 2.5257 & -0.478\\
6361.74569 & 4.4506 & 0.771\\
6439.07500 & 2.5257 & 0.390\\
6449.80800 & 2.5213 & -0.502\\
6450.06707 & 4.4506 & -0.632\\
6462.56700 & 2.5230 & 0.262\\
6493.78100 & 2.5213 & -0.109\\
Ca 2&&\\
5001.47900 & 7.5051 & -0.507\\
5019.97100 & 7.5148 & -0.247\\
\end{supertabular}
\end{center}

\newpage
\begin{center}
\tablecaption{Line data for Mg}
\vspace{5mm}
\tablefirsthead{\hline
\hline
\noalign{\smallskip}
Wavelength in air [Å] & $E_\text{exc}$ [eV] & $\log gf$ \\
\noalign{\smallskip}
\hline
\noalign{\smallskip}}

\tablehead{\small\sl ...continued\\
\hline
\noalign{\smallskip}
Wavelength in air [Å] & $E_\text{exc}$ [eV] & $\log gf$\\
\noalign{\smallskip}
\hline
\noalign{\smallskip}}
\begin{supertabular}{ccc}
Mg 1 &&\\
4702.99090 & 4.3458 & -0.440\\
4730.02860 & 4.3458 & -2.347\\
5167.32160 & 2.7091 & -0.931\\
5172.68430 & 2.7116 & -0.450\\
5183.60420 & 2.7166 & -0.239\\
5345.97600 & 5.1078 & -2.620\\
5345.97600 & 5.1078 & -2.840\\
5345.97600 & 5.1078 & -3.310\\
5528.40470 & 4.3458 & -0.498\\
\end{supertabular}
\end{center}

\begin{center}
\topcaption{Line data for Ti}
\vspace{5mm}
\tablefirsthead{\hline
\hline
\noalign{\smallskip}
Wavelength in air [Å] & $E_\text{exc}$ [eV] & $\log gf$ \\
\noalign{\smallskip}
\hline
\noalign{\smallskip}}

\tablehead{\small\sl ...continued\\
\hline
\noalign{\smallskip}
Wavelength in air [Å] & $E_\text{exc}$ [eV] & $\log gf$\\
\noalign{\smallskip}
\hline
\noalign{\smallskip}}
\begin{supertabular}{ccc}
Ti 1&&\\
4798.53062 & 3.7090 & -2.466\\
4911.05882 & 3.4406 & -2.348\\
4911.43762 & 3.5810 & -1.346\\
5153.91131 & 3.1864 & -1.677\\
5188.64081 & 2.2674 & -3.025\\
5188.82287 & 2.7778 & -1.450\\
5226.59419 & 3.3232 & -3.460\\
Ti 2&&\\
4798.53131 & 1.0800 & -2.660\\
4805.09283 & 2.0613 & -0.960\\
4911.19441 & 3.1235 & -0.640\\
5129.15625 & 1.8917 & -1.340\\
5154.06808 & 1.5658 & -1.716\\
5185.90188 & 1.8927 & -1.410\\
5188.68713 & 1.5818 & -1.020\\
5211.53016 & 2.5903 & -1.410\\
5226.53899 & 1.5658 & -1.185\\
5381.02174 & 1.5658 & -1.970\\
5418.76784 & 1.5818 & -2.130\\
\end{supertabular}
\end{center}

\end{appendix}


% ...

\end{document}